\documentclass[11pt]{iopart}

\usepackage{graphicx}
\usepackage{epsf}
\usepackage{longtable}
\usepackage{hyperref}
\usepackage{graphics,epsfig,placeins,subfigure,wrapfig}

\usepackage[usenames]{color}
\usepackage{soul}
\usepackage{amsfonts}
\usepackage[mathscr]{euscript}
\usepackage{amssymb}

\def\eadnew#1#2{\address{#2 E-mail: \mailto{#1}}}

\begin{document}

\title{Ideal Magnetohydrodynamics with Radiative Terms: Energy Conditions}

\author{Oscar M. Pimentel $^{1,*}$, 
F. D. Lora-Clavijo$^{1,\dag}$
Guillermo A. Gonz\'alez$^{1,\ddag}$,}
\address{$^1$ Grupo de Investigaci\'on en Relatividad y Gravitaci\'on, Escuela de F\'isica, Universidad Industrial de Santander, A. A. 678,
Bucaramanga 680002, Colombia.}
\eadnew{oscar.pimentel@correo.uis.edu.co}{$^{*}$}
\eadnew{fadulora@uis.edu.co}{$^{\dag}$}
\eadnew{guillermo.gonzalez@saber.uis.edu.co}{$^{\ddag}$}


\begin{abstract}
Nowadays, the magnetic and radiation fields are very important to understand the matter accretion into compact objects, the dynamics of binary systems, the equilibrium configurations of neutron stars, the photon diffusion, etc. The energy and the momentum associated to these fields, along with the matter one, need to satisfy some conditions that guarantee an appropriate physical behavior of the source and its gravitational field. Based on this fact, we present the energy conditions for a perfect fluid with magnetic and radiation field, in which the radiation part of the energy-momentum tensor is assumed to be approximately isotropic, in accordance with the optically thick regime. In order to find these conditions, the stress tensor of the system is written in an orthonormal basis in which it becomes diagonal, and the energy conditions are computed through contractions of the energy-momentum tensor with the four velocity vector of an arbitrary observer. Finally, the conditions for a magnetized fluid are presented as a particular case in which the radiation contribution is zero.
\end{abstract}

\section{Introduction}

It is usually expected that any solution to the Einstein field equations, describing realistic sources of gravitational field, has to satisfy the energy conditions \cite{1973lsss.book.....H}. These conditions for a perfect fluid and for a viscous fluid have been widely used in some astrophysical and theoretical contexts \cite{2010IJMPD..19.1783V, Gonzalez:2015qzt, 2013EPJP..128..123S}. Nevertheless, recent researches are focused on the magnetic and radiation fields contributions to the energy-momentum tensor, because they allow describing a wider variety of astrophysical scenarios; for instance, the effect of a toroidal magnetic field on the equilibrium configurations of rotating neutron stars was treated in \cite{2012MNRAS.427.3406F}. The structure of neutron stars when strong magnetic fields are presented is studied in \cite{Chatterjee:2014qsa}. Additionally, recent calculations have demonstrated that the binary neutron stars mergers, using magnetohydrodynamic simulations in full general relativity, produce relativistic jets and strong magnetic fields, which serve as a central engine for SGRBs \cite{2011ApJ...732L...6R, 2016arXiv160402455R}.  On the other hand, in some astrophysical systems the radiation is dynamically important, as in the case of the collapse of massive stars to black holes \cite{Farris:2008fe}, or the core-collapse supernovae, whose structure and dynamics is determined by the neutrino transport \cite{O'Connor:2014rva}. Additionally, the magnetic and radiation fields are key ingredients for giving a realistic description of the accretion of matter into a black hole. It was showed in \cite{2012MNRAS.426.1928D} that the radiative cooling processes should be considered for some accretion rates to reproduce the true dynamics of the disk. Finally, the optically thick limit for accretion disks is introduced numerically to study the Bondi \cite{Fragile:2012bs} and Bondi-Hoyle accretion \cite{Zanotti:2011mb} to black holes. 

On the other hand, one problem the relativistic Eulerian magnetohydrodynamic codes have to face is the implementation of an atmosphere to avoid the presence of unphysical results when the rest mass density is too small, that is, to avoid the presence of vacuum regions in the numerical domain. In the practice, what the different authors do is fix the value of the density $\rho$ and the velocity $v^{i}$ to $\rho=\rho_{atm}$ and $v^{i}=0$, once $\rho \le \rho_{atm}$ \cite{2007CQGra..24S.235G}, or set to zero the initial value of magnetic field, instead of the velocity, in the low density regions \cite{Shibata:2005gp}, and additionally,  impose constraints on the conservative variables to reduce posible numerical errors that can drive to unphysical results \cite{Etienne:2011re}. The correct values of $\rho_{atm}$, in principle, only depends on the specific problem under consideration; the only requirement is that the dynamics of the system should not be affected by the presence of such atmosphere \cite{2013rehy.book.....R}, which is usually treated as a low-density fluid governed by the same equation of state of the bulk matter \cite{Baiotti:2004wn, Galeazzi:2013mia}. Now, when the fluid has magnetic and radiation fields, the energy conditions may represent useful constraints on the physical variables, that motivate the atmospheres and the possible equations of state for that kind of source. For all the reasons mentioned, it becomes important to find the energy conditions for a perfect fluid with magnetic and radiation fields.

The energy conditions for a viscous fluid were calculated in \cite{2004rtmb.book.....P}. Nevertheless, when the heat flux vector of the fluid is different from zero, the procedure to compute the energy conditions becomes more difficult. In 1988 Kolassis, Santos, and Tsoubelis found the conditions for the energy-momentum tensor of a viscous fluid with heat flux, written in the Eckart approximation of small velocity gradients \cite{0264-9381-5-10-011}. To do it, the authors diagonalized the energy-momentum tensor and wrote its energy conditions in terms of the corresponding eigenvalues \cite{1973lsss.book.....H}. With this method, it was necessary to deal with a fourth order polynomial equation, so additional assumptions were required in order to find the eigenvalues. Recently, the energy conditions for the same dissipative fluid were found in \cite{2016arXiv160401318P}, without considering any approximation on the velocity gradients. The used method leave the energy conditions free from restrictions and consists in diagonalize the spatial part of the energy-momentum tensor, compute the physical quantities as measured by an arbitrary observer, and finally, to write the conditions so that they are independent of the observer. Now, in this work we are going to find the energy conditions for a perfect fluid with magnetic and radiation field by using the same algebraic procedure developed in \cite{2016arXiv160401318P}, that leaves the conditions independent from the 4-velocity of the observer.

The organization of this work is the following: We present in section \ref{sec:EMT} the energy-momentum tensor for a perfect fluid with magnetic and radiation fields, for which we are going to compute the energy conditions. We also discuss briefly the assumptions we made for each field and its physical motivation. In section \ref{sec:EC}, we compute the energy density and the isotropic pressure of the total system, because in terms of this quantities the energy-momentum tensor can be written in a similar fashion as the one of a viscous fluid with heat flux. Following the procedure presented in \cite{2016arXiv160401318P}, we find the energy conditions for the gravitational system consisting of a magnetized fluid with radiation field. As a particular case, we suppose the radiation contribution equals to zero in order to find the energy conditions for a perfect fluid with magnetic field. Finally, in section \ref{sec:conclusions} we discuss about the main results of this work. In this work, we will work with the signature $(-,+,+,+)$, and with units in which the gravitational constant $G$ and the speed of light $c$ are equal to one.

\section{The Energy-Momentum tensor for Relativistic Magneto-Fluids with Radiation Field}
\label{sec:EMT}

In this section we are going to present the energy-momentum tensor for a perfect fluid with magnetic and radiation fields. This tensor can be splitted into three terms,
\begin{equation}
\label{energy-momentum1}
T^{\alpha\beta}=T^{\alpha\beta}_{f}+T^{\alpha\beta}_{m}+T^{\alpha\beta}_{r},
\end{equation}
where $T^{\alpha\beta}_{f}$, $T^{\alpha\beta}_{m}$ and $T^{\alpha\beta}_{r}$ are the energy-momentum tensors of the fluid, the magnetic field, and the radiation field, respectively. Now, we will introduce each tensor by separately, presenting the approximations and assumptions for each contribution.
\begin{enumerate}
\item[\em{ a.}] {\em The fluid contribution}
\end{enumerate}
We are going to compute the energy conditions for a non-dissipative matter field in which the energy fluxes are zero and there are not anisotropic stresses. In this case, the energy-momentum tensor of the fluid can be written as
\begin{equation}
\label{m_fluid}
T^{\alpha\beta}_{f}=(\rho+p) u^{\alpha}u^{\beta}+pg^{\alpha\beta},
\end{equation} 
where $u^{\alpha}$ is the four-velocity vector of the fluid, $\rho$ is the energy density in the comoving frame, and $p$ is the thermal pressure of the fluid, also measured by the comoving observer. A fluid described by (\ref{m_fluid}) is commonly know as {\em perfect fluid}, and is widely used to model the average properties of isolated rotating relativistic stars \cite{2003LRR.....6....3S, 1992RSPTA.340..391F}, and to study the fluid dynamics around compact objects \cite{2015ApJS..219...30L, 2016arXiv160504176C}.

\begin{enumerate}
\item[\em{ b.}] {\em The magnetic field contribution}
\end{enumerate}
The energy-momentum tensor of the electromagnetic field is given by the following expression,
\begin{equation}
\label{energy-electromagnetic}
T^{\alpha\beta}_{em}=\frac{1}{4\pi}\left(F^{\alpha}_{~\gamma}F^{\beta\gamma}-\frac{1}{4}F_{\gamma\delta}F^{\gamma\delta}g^{\alpha\beta}\right),
\end{equation}
where $F_{\alpha\beta}$ is the Faraday tensor. This tensor can be decomposed in terms of the electric field $E^{\alpha}$, and the magnetic field, $B^{\alpha}$, measured by the comoving observer, as 
\begin{equation}
\label{electric_magnetic}
F^{\alpha\beta}=E^{\alpha}u^{\beta}-E^{\beta}u^{\alpha}+\frac{1}{2}\epsilon^{\alpha\beta\mu\nu}(u_{\mu}B_{\nu}-u_{\nu}B_{\mu})
\end{equation}
where $\epsilon^{\alpha\beta\gamma\delta}$ is the Levi-Civita tensor. 

On the other hand, according to the Ohm's law, $j^{\alpha}=\sigma E^{\alpha}$, where the conduction current, $j^{\alpha}$, is related to the electric field through the constitutive relation, $j^{\alpha}=\sigma E^{\alpha}$,
where $\sigma$ is the conductivity. Now, if we suppose that the material is a perfect conductor, then $\sigma\rightarrow \infty$, and the only way to have a finite conduction current is that $E^{\alpha}=0$. This approximation is the basis for the ideal magnetohydrodynamics, which is very useful to describe highly conducting astrophysical fluids where the effect of the magnetic field cannot be neglected, such as neutron stars, accretion fluids, magnetized winds, etc \cite{1989rfmw.book.....A}. As a consequence of this approximation, $F^{\alpha\beta}u_{\alpha}=0$; so it is possible to write the energy-momentum tensor (\ref{energy-electromagnetic}) as \cite{1989rfmw.book.....A, Lora-Clavijo:2014bba}
\begin{equation}
\label{mag_fluid}
T^{\alpha\beta}_{m}=|b|^{2}u^{\alpha}u^{\beta}+\frac{1}{2}|b|^{2}g^{\alpha\beta}-b^{\alpha}b^{\beta},
\end{equation}
where $|b|^{2}=b^{\alpha}b_{\alpha}$ and $b^{\alpha}=B^{\alpha}/\sqrt{4\pi}$. The 4-vector $b^{\alpha}$ is spacelike and satisfies the property $b^{\alpha}u_{\alpha}=0$. The tensor $T^{\alpha\beta}_{m}$ describes the energy and the momentum of the magnetic field in a well-conductor fluid.

\begin{enumerate}
\item[\em{ c.}] {\em The radiation field contribution}
\end{enumerate}

The last term in the right hand side of (\ref{energy-momentum1}), which describes the energy and the momentum of the radiation field, is given by \cite{1984oup..book.....M}
\begin{equation}
\label{Trad_definition}
T^{\alpha\beta}_{r}=\int I_{\nu}N^{\alpha}N^{\beta}d\nu d\Omega,
\end{equation}
where $I_{\nu}=I(x^{\alpha};N^{\alpha},\nu)$ is the specific intensity of the radiation at position $x^{\alpha}$ with frequency $\nu$ and moving in the direction $N^{\alpha}\equiv p^{\alpha}/h_{P}\nu$, $p^{\alpha}$ is the photon four-momentum, $h_{P}$ is the Planck constant, and $d\Omega$ is the differential of solid angle; $\nu$, $I_{\nu}$ and $d\Omega$ are measured in a comoving frame. Moreover, since the photon propagation direction becomes $N^{\hat{\alpha}}=(1,N^{\hat{i}})$, $\hat{i}=1,2,3$, in this comoving frame \cite{Farris:2008fe}, we can define the radiation moments \cite{1984oup..book.....M}, {\em i.e.}, the radiation energy density $T_{r}^{\hat{0}\hat{0}}=E_{r}$, the radiation flux $T_{r}^{\hat{0}\hat{i}}=F^{\hat{i}}_{r}$, and the radiation stress tensor $T_{r}^{\hat{i}\hat{j}}=P^{\hat{i}\hat{j}}_{r}$, as
\begin{equation}
E_{r}=\int I_{\nu}d\nu d\Omega, ~~~F^{\hat{i}}_{r}=\int I_{\nu}N^{\hat{i}}d\nu d\Omega, ~~~P^{\hat{i}\hat{j}}_{r}=\int I_{\nu}N^{\hat{i}}N^{\hat{j}}d\nu d\Omega.
\end{equation}

Now, many current numerical codes are devoted to describe systems such as relativistic stars or high density fluids in which the photons free paths are small, and therefore they interact with the material and diffuse through it. As a consequence, the radiation is trapped in the interior of the compact object and $I_{\nu}$ becomes isotropic, so that $I_{\nu}=I(x^{\alpha};\nu)$ \cite{1984oup..book.....M}. Under this limit of isotropy, widely known as the optically thick regime, the radiation stress tensor becomes $P^{\hat{i}\hat{j}}_{r}=P_{r}\delta^{\hat{i}\hat{j}}$,
with $P_{r}=E_{r}/3$, and the radiation flux vector vanishes \cite{Farris:2008fe}. Nevertheless, allowing $F^{\hat{i}}_{r}$ to have small non-zero values, we introduce a small anisotropy but retain the isotropy of the radiation stress tensor, so the radiation field is only approximately isotropic. Henceforth, we will concern only on gravitational systems in which the optically thick regime is approximately valid. The energy-momentum tensor of the radiation field in such a systems can be written in a covariant form as
\begin{equation}
\label{emt_raditaion_OTL}
T^{\alpha\beta}_{r}=(E_{r}+P_{r})u^{\alpha}u^{\beta}+F_{r}^{\alpha}u^{\beta}+F_{r}^{\beta}u^{\alpha}+P_{r}g^{\alpha
\beta},
\end{equation} 
where $F_{r}^{\alpha}$ is the radiation flux four-vector which is defined as
\begin{equation}
F_{r}^{\alpha}=h^{\alpha}_{~\beta}\int I_{\nu}N^{\beta}d\nu d\Omega,
\end{equation}
and $h^{\alpha\beta}=g^{\alpha\beta}+u^{\alpha}u^{\beta}$ is the induced metric of the normal space to the 4-velocity $u^{\alpha}$, so it is easy to show that $u_{\alpha}F_{r}^{\alpha}=0$ .

Finally, replacing (\ref{m_fluid}), (\ref{mag_fluid}), and (\ref{emt_raditaion_OTL}) in (\ref{energy-momentum1}) we obtain \cite{Farris:2008fe}
\begin{eqnarray}
\label{total_energy_momentum}
T^{\alpha\beta}&=&(\rho+p+|b|^{2}+E_{r}+P_{r})u^{\alpha}u^{\beta}+\left(p+P_{r}+\frac{1}{2}|b|^{2}\right)g^{\alpha\beta}\nonumber\\
&&+F_{r}^{\alpha}u^{\beta}+F_{r}^{\beta}u^{\alpha}-b^{\alpha}b^{\beta},
\end{eqnarray}
which is the total energy-momentum tensor for a perfect magneto-fluid with a radiation field.

\section{The Energy Conditions}
\label{sec:EC}

Now, we are interested in finding the energy conditions for a fluid described by (\ref{total_energy_momentum}). These conditions are restrictions on the energy-momentum tensor that ensure the solutions to the Einstein equations to describe realistic gravitational systems. We can briefly formulate the energy conditions as \cite{2004rtmb.book.....P, 2004sgig.book.....C}
\begin{itemize}
\item[{\em (a)}] {\em The weak energy condition:} The energy density, $\epsilon$, measured by an arbitrary observer, defined by its four-velocity vector $W^{\alpha}$, must be positive, {\em i.e.},
\begin{equation}
\label{WEC}
\epsilon=T_{\alpha\beta}W^{\alpha}W^{\beta}\geq 0.
\end{equation}
\item[{\em (b)}] {\em The strong energy condition:} Any timelike or null congruence of geodesics must be convergent. In other words, the gravitational field must be attractive. This condition is satisfied if the following inequality holds for any arbitrary observer \cite{1973lsss.book.....H},
\begin{equation}
\label{SEC}
\mu=R_{\alpha\beta}W^{\alpha}W^{\beta}\geq 0.
\end{equation}
Now, by using the Einstein equations, the last expression becomes
\begin{equation}
T_{\alpha\beta}W^{\alpha}W^{\beta}\geq-\frac{1}{2}T,
\label{SEC_1}
\end{equation}
so we can say that the stress of the fluid can not be much larger than its energy density.
\item[{\em (c)}] {\em The dominant energy condition:} The energy flux density measured by an arbitrary observer,
\begin{equation}
\label{energy_flux_vector}
S^{\alpha}=-T^{\alpha\beta}W_{\beta},
\end{equation}
must be a future oriented timelike or null vector. This means that
\begin{eqnarray}
S^{0}>0,\label{DEC1}\\
S^{\alpha}S_{\alpha}\geq 0.\label{DEC2}
\end{eqnarray}
\end{itemize}
This condition can be physically interpreted as saying that the matter and energy can not flow faster than light. We can see that the dominant energy condition contains the weak one, but all the three conditions are mathematically independent \cite{1984ucp..book.....W}. Therefore, the aim of this section is to write this conditions in terms of the energy density and pressure of the fluid, the energy density and pressure of the radiation, the radiation flux vector, and the magnetic field.

There are two ways for addressing the problem. The first one is to rewrite the energy conditions in terms of the eigenvalues of $T^{\alpha\beta}$ \cite{1973lsss.book.....H, 0264-9381-5-10-011}. Nevertheless, computing the eigenvalues of (\ref{total_energy_momentum}) is a difficult task because we need to deal with a fourth-order polynomial, so it becomes necessary to impose additional restrictions on the energy-momentum tensor in order to find solutions. The other way for finding the energy conditions is to define an orthonormal basis in which the spatial part of $T^{\alpha\beta}$ (the one associated with the normal and tangential stresses) becomes diagonal. In this basis, the four-velocity $W^{\alpha}$ has arbitrary components, and therefore, the energy conditions are written in terms of these components (cf. \ref{WEC} - \ref{DEC2}). Then, with the aim of ensuring the invariance of the energy conditions, it is necessary to apply algebraic procedures to decouple them from the components of $W^{\alpha}$. This last method was applied in \cite{2016arXiv160401318P} in order to find the energy conditions for a viscous fluid with heat flux.

In this paper, we are going to use the same method used in \cite{2016arXiv160401318P}, but now with the energy-momentum tensor of a perfect magneto-fluid with radiation field (\ref{total_energy_momentum}). To do it, we start by computing the total energy of the system, $\mathscr{E}$, measured by a comoving observer, 
\begin{equation}
\label{energy}
\mathscr{E}=T^{\alpha\beta}u_{\alpha}u_{\beta}=\rho+\frac{1}{2}|b|^{2}+E_{r}.
\end{equation}
We also compute the isotropic pressure \cite{2000ifd..book.....B}, which is written via the spatial trace of the energy-momentum tensor as
\begin{equation}
\label{isotropic_pressure}
\hat{\mathscr{P}}=\frac{1}{3}h_{\alpha\beta}T^{\alpha\beta}=p+P_{r}+\frac{1}{6}|b|^{2}.
\end{equation}
Now, in terms of $\mathscr{E}$ and $\hat{\mathscr{P}}$ the energy-momentum tensor (\ref{total_energy_momentum}) takes the form
\begin{equation}
\label{total_T_v1}
T^{\alpha\beta}=(\mathscr{E}+\hat{\mathscr{P}})u^{\alpha}u^{\beta}+\hat{\mathscr{P}}
g^{\alpha\beta}+F_{r}^{\alpha}u^{\beta}+F_{r}^{\beta}u^{\alpha}+\overline{\Pi}^{\alpha\beta},
\end{equation}
where we have defined the deviatoric tensor
\begin{equation}
\label{deviatoric_stress}
\overline{\Pi}^{\alpha\beta}=\frac{1}{3}|b|^{2}h^{\alpha\beta}-b^{\alpha}b^{\beta}.
\end{equation}
It is important to note that $\overline{\Pi}^{\alpha\beta}$ satisfies the same properties of the deviatoric stress tensor from relativistic fluid dynamics \cite{2000ifd..book.....B}: it is traceless and its projection along $u_{\alpha}$ is zero, so
\begin{equation}
\label{DS_Property}
g_{\alpha\beta}\overline{\Pi}^{\alpha\beta}=\overline{\Pi}^{\alpha}_{~\alpha}=0,\hspace{10mm} 
\overline{\Pi}^{\alpha\beta}u_{\alpha}=\overline{\Pi}^{\alpha\beta}u_{\beta}=0,
\end{equation}
respectively. Finally, if we introduce the stress tensor
\begin{equation}
\label{stress_tensor}
\mathscr{S}^{\alpha\beta}=\hat{\mathscr{P}}h^{\alpha\beta}+\overline{\Pi}^{\alpha\beta},
\end{equation}
then (\ref{total_T_v1}) reduces to
\begin{equation}
\label{T_final_v}
T^{\alpha\beta}=\mathscr{E}u^{\alpha}u^{\beta}+F_{r}^{\alpha}u^{\beta}+F_{r}^{\beta}u^{\alpha}+\mathscr{S}^{\alpha\beta},
\end{equation}
which has the same form as the one of the energy-momentum tensor for a viscous fluid with heat flux. This make sense since the term $b^{\alpha}b^{\beta}$ may introduce tangential stresses due to the magnetic field, in a similar manner as the deviatoric stress tensor does for a viscous fluid. Additionally, the radiation flux vector, $F_{r}^{\alpha}$, plays an analogous roll to the one of the heat flux vector.

The tensor $\mathscr{S}^{\alpha\beta}$ is the spatial part of $T^{\alpha\beta}$; therefore, it describes the normal and tangential stresses due to the fluid, the magnetic field, and the radiation field. We can find an orthonormal basis of eigenvectors $\{X^{\alpha}, Y^{\alpha}, Z^{\alpha}\}$ in which $\mathscr{S}^{\alpha\beta}$ takes the diagonal form,
\begin{equation}
\label{stress_eigenvectors}
\mathscr{S}^{\alpha\beta}=\mathscr{P}_{1}X^{\alpha}X^{\beta}+\mathscr{P}_{2}Y^{\alpha}Y^{\beta}+
\mathscr{P}_{3}Z^{\alpha}Z^{\beta},
\end{equation}
where, $\mathscr{P}_{i}$, $i=1,2,3,$ are the eigenvalues. In this basis of eigenvectors, the radiation flux and the magnetic field have arbitrary components, so,
\begin{eqnarray}
F^{\alpha}_{r}&=&F_{r_{1}}X^{\alpha}+F_{r_{2}}Y^{\alpha}+F_{r_{3}}Z^{\alpha}, \label{radiation_eigenvectors}\\
b^{\alpha}&=&\mathfrak{b}_{1}X^{\alpha}+\mathfrak{b}_{2}Y^{\alpha}+\mathfrak{b}_{3}Z^{\alpha},
\end{eqnarray}
in such a way that $(F_{r})^{2}=(F_{r_{1}})^{2}+(F_{r_{2}})^{2}+(F_{r_{3}})^{2}$ and $|b|^{2}=b_{\alpha}b^{\alpha}=(\mathfrak{b}_{1})^{2}+(\mathfrak{b}_{2})^{2}+(\mathfrak{b}_{3})^{2}$. Additionally, in the orthonormal tetrad $\{u^{\alpha}, X^{\alpha}, Y^{\alpha}, Z^{\alpha}\}$, the four-velocity vector of the arbitrary observer takes the form
\begin{equation}
\label{velocity_observer}
W^{\alpha}=\gamma u^{\alpha}+A_{1}X^{\alpha}+A_{2}Y^{\alpha}+A_{3}Z^{\alpha},
\end{equation}
where $\gamma$ and $A_{i}$, $i=1,2,3,$ satisfy the relation $\gamma=\sqrt{1+A^{2}}\geq 1$, with $A^{2}=A_{1}^{2}+A_{2}^{2}+A_{3}^{2}$. Finally, the eigenvalues, $\mathscr{P}_{i}$ can be computed as follows:
\begin{eqnarray}
\mathscr{P}_{1}&=&\mathscr{S}^{\alpha\beta}X_{\alpha}X_{\beta}=p+P_{r}+\frac{1}{2}|b|^{2}-\mathfrak{b}_{1}^{2},\label{iso1}\\
\mathscr{P}_{2}&=&\mathscr{S}^{\alpha\beta}Y_{\alpha}Y_{\beta}=p+P_{r}+\frac{1}{2}|b|^{2}-\mathfrak{b}_{2}^{2},\label{iso2}\\
\mathscr{P}_{3}&=&\mathscr{S}^{\alpha\beta}Z_{\alpha}Z_{\beta}=p+P_{r}+\frac{1}{2}|b|^{2}-\mathfrak{b}_{3}^{2},\label{iso3}
\end{eqnarray}
where the different magnetic components are the responsible for the anisotropy of $\mathscr{S}^{\alpha\beta}$, because in general, $\mathscr{P}_{1}\neq\mathscr{P}_{2}\neq\mathscr{P}_{3}$.

The following step is to use (\ref{T_final_v}) and (\ref{velocity_observer}) to write the inequalities (\ref{WEC} - \ref{DEC2}) and try to develop an algebraic procedure to decouple the energy conditions from the components of $W^{\alpha}$. Nevertheless, this algebraic procedure will be the same as the one used in \cite{2016arXiv160401318P} to find the energy conditions for a viscous fluid with heat flux. Hence, we can state that the energy conditions for the energy-momentum tensor presented in (\ref{T_final_v}) are (see Appendix A)
\begin{eqnarray}
\mathscr{E} \geq 2 F_{r} ,\label{e11} \\
\mathscr{E} + \mathscr{P}_{i} \geq 2 F_{r} ,\qquad i = 1,2,3, \label{e22} \\
\mathscr{E} + \mathscr{P}_{1} + \mathscr{P}_{2} + \mathscr{P}_{3} \geq 4 F_{r} \label{e33}.\\
\mathscr{E}^2 \geq \mathscr{P}_{i}^2 + F_{r}^2 + 2 (\mathscr{E} +  \mathscr{P}_{1} +  \mathscr{P}_{2} +  \mathscr{P}_{3}) F_{r}, \qquad i = 1,2,3. \label{e55}
\end{eqnarray}
Therefore, using (\ref{energy}), (\ref{isotropic_pressure}), and (\ref{iso1}-\ref{iso3}), the energy conditions for a perfect magneto-fluid with radiation field are
\begin{eqnarray}
\rho+\frac{1}{2}|b|^{2}+E_{r} \geq 2 F_{r} ,\label{e111} 
\end{eqnarray}
\begin{eqnarray}
\rho+|b|^{2}+E_{r}+p+P_{r}-\mathfrak{b}_{i}^{2} \geq 2 F_{r} ,\quad i = 1,2,3, \label{e222} 
\end{eqnarray}
\begin{eqnarray}
\rho+|b|^{2}+E_{r}+3p+3P_{r} \geq 4 F_{r} \label{e333}.
\end{eqnarray}
\begin{eqnarray}
(\rho+\frac{1}{2}|b|^{2}+E_{r})^2 \geq && \left(p+P_{r}+\frac{1}{2}|b|^{2}-\mathfrak{b}_{i}^{2}\right)^2 + F_{r}^2\nonumber\\&&+2(\rho+|b|^{2}+E_{r}+3p+3P_{r}) F_{r}, \quad i = 1,2,3. \label{e555}
\end{eqnarray}
With the inequalities (\ref{e111}), (\ref{e222}), and (\ref{e333}) the weak energy condition and the strong energy condition are both satisfied; while the dominant energy condition is equivalent to (\ref{e555}).

As a particular case, when $E_{r}=0$, $P_{r}=0$ and $F_{r}^{\alpha}=0$, the radiation field vanishes, and (\ref{e111}-\ref{e555}) reduce to 
\begin{eqnarray}
\rho+\frac{1}{2}|b|^{2}\geq 0 ,\label{e1} \\
\rho+p+|b|^{2}-\mathfrak{b}_{i}^{2} \geq 0 ,\quad i = 1,2,3, \label{e2} \\
\rho+3p+|b|^{2} \geq 0 \label{e3}, \\
\rho \geq |p-\mathfrak{b}_{i}^{2}| , \quad i = 1,2,3. \label{e5}
\end{eqnarray}
These energy conditions are also of astrophysical and numerical interest because they may be applied to the energy-momentum tensor of a perfect fluid with an arbitrary magnetic field.

\section{Conclusion}
\label{sec:conclusions}

We have obtained the energy conditions for a perfect fluid with magnetic and radiation fields, whose physical description is the main objective of the {\em general relativistic radiation magneto-hydrodynamics}. We first presented the energy-momentum tensor of the total system (\ref{total_energy_momentum}), consisting of the fluid, the magnetic field, and the radiation field. With this tensor, we computed the energy density of the system, by projecting (\ref{total_energy_momentum}) along the four-velocity vector of the fluid, and the isotropic pressure, which is defined through the spatial trace of (\ref{total_energy_momentum}). By writing the energy-momentum tensor in terms of these quantities, we realized that the magnetic field introduces an anisotropic term, $\overline{\Pi}^{\alpha\beta}$, to the stress tensor of the fluid (\ref{stress_tensor}), in the same way as the deviatoric stress tensor does for a viscous fluid.

By writing the four-velocity vector of the arbitrary observer $W^{\alpha}$ in the comoving tetrad, where the stress tensor (\ref{stress_tensor}) is diagonal, we can compute the energy density $\epsilon$, the energy flux density $S^{\alpha}$, and the scalar $\mu$, as measured by this observer. Then, by applying the same algebraic procedure presented in \cite{2016arXiv160401318P}, we showed that the weak, and strong energy conditions are satisfied if (\ref{e111}-\ref{e333}) are simultaneously satisfied. We also showed that the dominant energy condition, for the system of interest, is equivalent to the inequality (\ref{e555}). As a particular case, when $E_{r}=0$, $P_{r}=0$, and $F_{r}^{\alpha}=0$, the radiation field vanishes, and the energy conditions reduce to those presented in (\ref{e1}-\ref{e5}), and correspond to a perfect fluid with magnetic field. These conditions are useful since many numerical simulations are carried out using this fluid as a test fluid in a curved spacetime background, or as a source of gravitational field to model the average properties of neutron stars and white dwarfs.

Finally, it is worth mentioning that the procedure to compute the energy conditions from the energy-momentum tensor (\ref{total_energy_momentum}) is quite general, and can be applied to any $T^{\alpha\beta}$. In particular, it is possible to consider fluids in which the specific intensity of the radiation, $I_{\nu}$, is not isotropic, and therefore the radiation stress tensor can be non-diagonal. Nevertheless, we have worked with this approximation, first of all with the aim of simplifying the calculations and the results, and secondly because the optically thick limit is used in most of the codes designed to study the radiation process associated with the dynamics of accretion disks around compact objects in the frame of the general relativity.

\section*{Acknowledgments}
\label{sec:conclusions}

O. M. P. wants to thanks the financial support from COLCIENCIAS and Universidad Industrial de Santander. F.D.L-C gratefully acknowledges the financial support from Universidad Industrial de Santander under grant number 1822. G. A. G. was supported in part by VIE-UIS, under Grants No. 1347 and No. 1838, and by COLCIENCIAS, Colombia, under Grant No. 8840.

\appendix

\section{Derivation of the energy conditions}

In this appendix we are going to present a brief summary of the algebraic procedure to compute the energy conditions (\ref{e11} - \ref{e55}) for a perfect fluid with magnetic and radiation fields. Nevertheless, this procedure is explained with more detail in \cite{2016arXiv160401318P}.

To start, we use the energy-momentum tensor (\ref{T_final_v}) and the four-velocity vector of the arbitrary observer (\ref{velocity_observer}), to write the energy density, $\epsilon$, and the term $\mu=R_{\alpha\beta}W^{\alpha}W^{\beta}$, as
\begin{eqnarray}
\epsilon &=& \mathscr{E} \gamma^2 + \sum_{i=1}^{3} \mathscr{P}_{i} A_i^2- 2 \gamma ( {\bf F_{r}} \cdot {\bf A} ),\label{eq1}\\
\mu &=& \frac{1}{2}\left(\mathscr{E} + \sum_{i=1}^{3} \mathscr{P}_i\right) +
\sum_{i=1}^{3}(\mathscr{E} + \mathscr{P}_i) A_i^2 - 2 \gamma ( {\bf F_{r}} \cdot {\bf A}
),
\end {eqnarray}
respectively. Now, using the relations, $\gamma^{2}=1-A^{2}$ and $\gamma\geq A$, and taking into account that $-F_{r}A \leq {\bf F_{r}} \cdot {\bf A} \leq F_{r}A$, we can conclude, after some calculations, that 
\begin{eqnarray}
\epsilon \geq (\mathscr{E} - 2 F_{r}) + \sum_{i=1}^{3}(\mathscr{E} + \mathscr{P}_i-2 F_{r}) A_i^2,\\
\mu \geq \frac{\mathscr{E} + 3 \hat{\mathscr{P}} - 4 F_{r}}{2} + \sum_{i=1}^{3}(\mathscr{E} + \mathscr{P}_i-2 F_{r})
A_i^2.
\end{eqnarray}
Therefore, the necessary and sufficient conditions to have $\epsilon\geq 0$, {\em i.e.} to satisfy the weak energy condition for all values of $A_{i}$ are 
\begin{eqnarray}
\mathscr{E} \geq 2 F_{r} , \label{condi1}\\
\mathscr{E} + \mathscr{P}_i \geq 2 F_{r}\label{condi2} ,
\end{eqnarray}
with $i=1,2,3$. Equivalently, the strong energy condition ($\mu\geq 0$) is satisfied for all values of $A_{i}$ if
\begin{eqnarray}
&& \mathscr{E} + 3 \hat{\mathscr{P}} \geq 4 F_{r} , \label{condi3} \\
&& \mathscr{E} + \mathscr{P}_i \geq 2 F_{r} \label{condi4},
\end{eqnarray}
where $i=1,2,3$.

On the other hand, with the aim of computing the dominant energy condition in terms of the physical quantities, we write the energy flux density, $S^{\alpha}$ in a comoving tetrad through the transformation $S_{(\mu)}=e_{(\mu)}^{\alpha}S_{\alpha}$, where $e_{(\mu)}^{\alpha}$ are orthonormal vectors. In this way, the metric tensor reduces to that of Minkowski and the calculations are easier. The condition (\ref{DEC1}) can be written as $S^{(0)} = \mathscr{E} \gamma - {\bf F_{r}} \cdot {\bf A} > (\mathscr{E} - F_{r}) \gamma$, and therefore, $S^{(0)}>0$ if $\mathscr{E} > F_{r}$.

Finally, the term $S^{\alpha}S_{\alpha}$ in the condition (\ref{DEC2}) becomes $S_{\alpha} S^{\alpha} = S_{(\mu)} S^{(\mu)} = -N(S)$ with $N(S)= (S^{(0)})^2 - (S^{(1)})^2 - (S^{(2)})^2 - (S^{(3)})^2$ and $S^{(i)}=F_{r_{i}}\gamma - \mathscr{P}_i A_i$. In this way, we can show that
\begin{equation}
N(S)=(\mathscr{E} \gamma - {\bf F_{r}} \cdot {\bf A})^2 - \sum_{i=1}^{3}(F_{r_{i}}
\gamma - \mathscr{P}_i A_i)^2.
\end{equation}
Then, expanding this expression, using the condition $\mathscr{E} > F_{r}$, and the fact that $F_{r}>0$, we can show that
\begin{eqnarray}
N(S)&\geq& [\mathscr{E}^2- F_{r}^2-2(\mathscr{E} + 3 \hat{\mathscr{P}})F_{r}]\nonumber
\\&+&\sum_{i=1}^{3}[\mathscr{E}^2-F_{r}^2-2(\mathscr{E}+3\hat{\mathscr{P}})F_{r}- \mathscr{P}_i^2]A_i^2.
\end{eqnarray}
Hence, in order to satisfy the dominant energy condition ($N(S)\geq 0$) for all values of $A_{i}$, then
\begin{eqnarray}
\mathscr{E}^2 &\geq& F_{r}^2 + 2 (\mathscr{E} + 3 \hat{\mathscr{P}}) F_{r}, \label{condi6} \\
\mathscr{E}^2 &\geq& F_{r_{i}}^2 + F_{r}^2 + 2 (\mathscr{E} + 3 \hat{\mathscr{P}}) F_{r} \label{condi7},
\end{eqnarray}
where $i=1,2,3$. We can see that the condition (\ref{condi7}) contains (\ref{condi6}) and therefore, all the energy conditions are satisfied if (\ref{condi1}), (\ref{condi2}), (\ref{condi3}) and (\ref{condi7}) are simultaneously satisfied.

\section*{References}




\end{document}